\pdfoutput=1
\documentclass[runningheads]{llncs}
\usepackage[T1]{fontenc}
\usepackage[ruled]{algorithm}
\usepackage{algorithmicx}
\usepackage{algpseudocode}
\usepackage{booktabs}
\usepackage{xcolor}
\usepackage{appendix}
\usepackage{listings}
\usepackage{amsmath}
\usepackage{stmaryrd}
\usepackage{multirow}
\usepackage{marvosym}
\usepackage{amssymb}
\usepackage{hyperref}
\usepackage{cite}
\hypersetup{colorlinks,
	linkcolor=blue,%
	citecolor=blue}
\usepackage{graphicx}
\UseRawInputEncoding

\let\oldReturn\Return
\renewcommand{\Return}{\State\oldReturn}
\makeatletter
\renewcommand*\env@matrix[1][\arraystretch]{%
  \edef\arraystretch{#1}%
  \hskip -\arraycolsep
  \let\@ifnextchar\new@ifnextchar
  \array{*\c@MaxMatrixCols c}}
\makeatother

\newcommand{\Braces}[1]{\ensuremath{\left\{#1\right\}}}

\newcommand{\Array}[1]{\ensuremath{\Braces{#1}}}
\newcommand{\SUM}[3]{\ensuremath{\displaystyle\sum_{#1}^{#2}{#3}}}

\renewcommand{\epsilon}{\varepsilon}

\newcommand{\VT}[1]{\ensuremath{\textbf{\lowercase{#1}}}}

\allowdisplaybreaks[4]

\newcommand{\RingRq}[1][q]{\ensuremath{\mathcal{R}_{#1}}}

\newcommand{\topcaption}{%
\setlength{\abovecaptionskip}{0pt}%
\setlength{\belowcaptionskip}{10pt}%
\caption}   
\AtBeginDocument{%
  }
\begin{document}
\begin{sloppypar}
\title{ESPM-D: Efficient Sparse Polynomial Multiplication for Dilithium on ARM Cortex-M4 and Apple M2}
\titlerunning{ESPM-D}

%
%
\author{Jieyu Zheng \and
Hong Zhang \and
Le Tian \and
Zhuo Zhang \and
Hanyu Wei \and
Zhiwei Chu \and Yafang Yang \and
Yunlei Zhao$^{\href{mailto:ylzhao@fudan.edu.cn}{\textrm{\Letter}}}$}
\authorrunning{J. Zheng et al.}
%
\institute{School of Computer Science and Technology, Fudan University, Shanghai, China
\email{\{jyzheng23, hongzhang22, tianl22, zhuozhang22, hywei22, zwchu23\}@m.fudan.edu.cn, \{18110240046, ylzhao\}@fudan.edu.cn}}
%
\maketitle              
\begin{abstract}

\textsf{Dilithium} is a lattice-based digital signature scheme standardized by the NIST post-quantum cryptography (PQC) project. In this study, we focus on developing efficient sparse polynomial multiplication implementations of \textsf{Dilithium} for ARM Cortex-M4 and Apple M2, which are both based on the ARM architecture. The ARM Cortex-M4 is commonly utilized in resource-constrained devices such as sensors. Conversely, the Apple M2 is typically found on mobile devices, emphasizing high performance and versatility. Accordingly, our optimization strategies differ between ARM Cortex-M4 and Apple M2. We prioritize optimizing stack usage for the former while enhancing computational efficiency for the latter. Our optimized sparse polynomial multiplication achieves significant speedups of up to 30\% on ARM Cortex-M4 and 55\% on Apple M2 compared to the state-of-the-art Number-Theoretic Transform (NTT) implementation. Additionally, we integrate the sparse polynomial multiplication with the infinity norm judgments in the \textsf{Dilithium} signing process, further enhancing signing efficiency. Our optimized implementation not only reduces stack usage by 10.8\%, 1.2\%, and 7.7\% in the signing procedure of \textsf{Dilithium2}, \textsf{Dilithium3}, and \textsf{Dilithium5}, respectively, but also enhances signing performance by 0.4\% to 0.8\% compared to the state-of-the-art ARM Cortex-M4 implementation. Furthermore, we optimize polynomial sampling, rounding functions, and polynomial packing and unpacking using ARM Cortex-M4 DSP instructions, resulting in a 0.4\%-3.2\% improvement in key generation and verification procedures. On the MacBook Air 2022, our \textsf{Dilithium} implementation achieves 10\% to 11\% speedups in the signing procedure. To the best of our knowledge, our work sets new performance records for \textsf{Dilithium} on both ARM Cortex-M4 and Apple M2 platforms.

\end{abstract}
\section{Introduction}
Digital signatures play a crucial role in ensuring the non-repudiation of messages sent by the sender. As of now, the RSA algorithm is widely used for most digital signatures \cite{1978A}. However, the landscape is evolving with the introduction of Shor's algorithm \cite{Shor1999Polynomial}, specifically designed for large number factorization. This algorithm poses a significant threat to RSA, as quantum computers can leverage Shor's algorithm to break it.

In response to the growing threat posed by quantum computers to public-key cryptography, the National Institute of Standards and Technology (NIST) has taken proactive measures by initiating a solicitation for post-quantum cryptography (PQC). Notably, on July 5, 2022, NIST announced the early selection of four PQC schemes, which include \textsf{Kyber} \cite{avanzi2017crystals}, \textsf{Falcon} \cite{fouque2018falcon}, \textsf{Dilithium} \cite{lyubashevsky2020crystals}, and \textsf{SPHINCS+} \cite{bernstein2019sphincs+}, as part of its PQC project standardization algorithm results. In this work, we focus on the \textsf{Dilithium} algorithm. It consists of three parameter sets referred to as \textsf{Dilithium2}, \textsf{Dilithium3}, and \textsf{Dilithium5}, respectively, where \textsf{Dilithium3} is the recommended parameter set. The impending shift into the post-quantum era poses unique challenges for micro-sized embedded devices, characterized by limited memory, low computational capabilities, and heightened susceptibility to security attacks. To address these challenges, researchers are actively exploring the feasibility of implementing PQC schemes on embedded platforms, with specific attention given to mainstream ARM Cortex-M4 processors.
\paragraph{Motivations. }
The recent study by \textsf{Dilithium} on ARM Cortex-M4 \cite{abdulrahman2022faster} introduced a specialized small-modulus NTT tailored for polynomial multiplication with a small modulus. This advancement aimed to accelerate the signature generation process for \textsf{Dilithium}. Huang et al. \cite{huang2024revisiting} revisited the work of \cite{abdulrahman2022faster} and improved \textsf{Keccak}'s performance on Cortex-M4. They pointed out that the small polynomial multiplication in \cite{zheng2023optimized} is unsuitable for the Cortex-M4 platform with a register width of 32 bits. In a related study, Becker et al.  \cite{2021Neon} demonstrated a vectorized NEON NTT implementation designed for the Apple M1 platform. Zheng et al. \cite{zheng2022parallel} presented a rapid implementation of small polynomial multiplication on the ARMv8 architecture, contributing to the enhanced performance of \textsf{Dilithium} on ARMv8. However, the implementation of sparse polynomial multiplication in \textsf{Dilithium} targeted ARM Cortex-M4 and Apple M2 has not been discussed. It's also interesting to investigate a sparse polynomial multiplication implementation that is suitable for platforms with low-width registers like ARM Cortex-M4. Therefore, we address the challenge stated in \cite{huang2024revisiting} about PSPM in \cite{zheng2022parallel} requiring larger bit-width registers. 

 
 \paragraph{Contributions. } In this paper, we select the ARM Cortex-M4 and the Apple M2 as our embedded platforms. While our primary aim is to optimize sparse polynomial implementation across both platforms, our approach differs based on their individual characteristics. The ARM Cortex-M4, endorsed by NIST for embedded development in PQC, necessitates a focus on stack usage optimization due to its constrained resources. In contrast, the Apple M2, renowned for its high-performance capabilities, directs our emphasis towards performance optimization. The contributions of this work are summarized as follows.
 \begin{itemize}

    \item We optimize the sparse polynomial multiplication proposed in the work \cite{zheng2022parallel}, introducing the methods to encode the challenge polynomial and implement branchless sparse polynomial multiplication. The optimized sparse polynomial multiplication is more suitable for implementation on embedded ARM platforms and outperforms the state-of-the-art small-modulus NTT implementation in \cite{huang2024revisiting}.

    \item We explore the possibilities of sparse polynomial multiplication parallelism in \textsf{Dilithium} with different parameters on ARM Cortex-M4 and Apple M2. We propose the use of 8-bit storage for secret polynomial vector coefficients, achieving 4-way parallelism on ARM Cortex-M4 by utilizing the Digital Signal Processor (DSP) instructions and 16-way parallelism on Apple M2 by utilizing the NEON 128-bit vector registers.

    \item We propose two versions of sparse polynomial multiplication implementation, wherein we mitigate the repeated accesses by strategically adjusting the branch order within the rejection sampling loop. Additionally, we minimize the loop jump counts by leveraging the \texttt{.rept} pseudo-instruction and optimizing the floating-point register cache. To further enhance implementation efficiency, we incorporate the  DSP instructions, facilitating the parallelization of both addition and subtraction processes.

    \item We implement the optimized sparse polynomial multiplication on  Apple M2. Our approach involves using the  ARM NEON Intrinsics for vectorized sparse polynomial multiplications. Furthermore, we combine the  sparse polynomial multiplication with the infinite norm check using parallel comparison functions in ARM NEON.

    \item We present the first optimized implementation using DSP instructions for functions:  polynomial sampling for polynomial vectors $\mathbf{s}_1$, $\mathbf{s}_2$, and $\mathbf{y}$, polynomial encoding and decoding. The optimized functions implementation improves key generation and verification by 0.4\% to 3.2\% on ARM Cortex-M4.

    \item Finally, our optimized sparse polynomial multiplication on ARM Cortex-M4 and Apple M2 brings improved performance for \textsf{Dilithium} and sets a new speed record for \textsf{Dilithium} on ARM Cortex-M4 and Apple M2.

\end{itemize}
It should be noted that our optimizations for sparse polynomial multiplication are not limited to
ARM Cortex-M4 and Apple M2. They can be extended to other lattice-based cryptographic schemes involving sparse polynomial multiplication and applied to ARM 32-bit and 64-bit processors based on ARMv7 and ARMv8 architectures.  We will open source our code later.


\paragraph{Related Work. }
Currently, multiple implementations of PQC schemes on ARM Cortex-M4 processors exist \cite{kannwischer2019faster, heinz2022first, paksoy2022faster, anastasova2022time, Chen2021, abdulrahman2022faster, greconici2021compact, chung2021ntt, Alkim2020, alkim2022multi}. This ongoing research and development initiative is pivotal in ensuring the adaptability of embedded devices to the evolving landscape of cryptographic security, especially considering the imminent quantum advancements. In a related context, Campos et al. \cite{2020LMS} contributed to the field by enhancing and comparing implementations of stateful hash-based signatures \textsf{LMS} and \textsf{XMSS} on the Cortex-M4 processor. Kannwischer et al. \cite{2019pqm4} introduced the \texttt{pqm4} testing and evaluation framework for PQC schemes, specifically on the \texttt{STM32F4Discovery} development board. Greconici et al. \cite{greconici2021compact} implemented the \textsf{Dilithium} signature scheme on ARM Cortex-M3 and ARM Cortex-M4, achieving performance improvement on both platforms. Guneysu et al. presented implementations of \textsf{GLP}, \textsf{BLISS}, and \textsf{Dilithium} on ARM Cortex-M4 \cite{guneysu2018evaluation}. Chou et al. \cite{CHES:ChoKanYan21} utilized a fast constant-time bitwise slicing $\mathbb{F}_{16}$ multiplication with 32 multiplications in 32 cycles. Kim et al. \cite{kim2022accelerating} utilized the ARMv8 NEON engine to accelerate FFT and NTT, achieving the first \textsf{Falcon} implementation on ARMv8. Nguyen et al. \cite{nguyen2023fast} set new speed records for \textsf{Falcon} signing and verification processes on ARMv8 processors. Zhao \cite{zhao2021efficient} et al. parallelized the core modules of \textsf{Kyber}, such as NTT and Inverse NTT (INTT), to achieve acceleration on ARMv8-A. Becker et al. \cite{2021Neon} achieved faster implementations of \textsf{Dilithium}, \textsf{Kyber}, and \textsf{Saber} on Cortex-A72 and Apple M1, holding the current record for the fastest \textsf{Dilithium} implementation on the ARMv8-A architecture.

\section{Preliminaries}
\subsection{Notation}
We adhere to the notation as outlined in \textsf{Dilithium} \cite{ducas2018crystals}. Let $n$ be a parameter representing a power of two, and $q$ be a prime satisfying $2n | (q-1)$. We denote $R$ as the polynomial ring $\mathbb{Z}[x] /\left(x^n+1\right)$, and $R_q$ as the polynomial ring $\mathbb{Z}_q[x] /\left(x^n+1\right)$. We denote the elements of $R_q$ by lower-case letters (e.g., $c$). We denote $B_\tau$ as the set of elements of $R$ that have $\tau$ coefficients that are either -1 or 1 and the rest are 0. We use bold upper-case letters ($\mathbf{A}$) to denote polynomial matrices and bold lower-case letters ($\mathbf{s}$) to represent polynomial vectors. In the NTT domain, elements are denoted with a hat symbol, for example, $\hat{c}, \hat{\mathbf{A}}, \hat{\mathbf{s}}$. We define $r^{\prime}=r \bmod ^{ \pm} q$ to be $r^{\prime}$ in the range $-\frac{q-1}{2} \leq r^{\prime} \leq \frac{q-1}{2}$. We denote $\|\cdot\|_{\infty}$ as the infinity form. For an element $w\in\mathbb{Z}_q$, $\|w\|_{\infty}=w \bmod ^{ \pm} q$. For a length-$m$ vector $\mathbf{w}$ with entries from $R_q$, $\|\mathbf{w}\|_{\infty}=\max _{0 \leq i<m}\|w[i]\|_{\infty}$. 

\subsection{Dilithium Signature Scheme}
\textsf{Dilithium} is a lattice-based signature scheme, and its security is derived from the underlying lattice problem, specifically the Module-Learning With Errors (MLWE) problem and Module Short Integer Solution (MSIS). The overall framework of \textsf{Dilithium} is inspired by the Fiat-Shamir transformation \cite{lyubashevsky2009fiat, lyubashevsky2012lattice}. During the signing process, signatures failing to meet the conditions are rejected, and the system regenerates new signatures until all conditions are satisfied. \textsf{Dilithium} contains three procedures: key generation (\textsf{KeyGen}), signing (\textsf{Sign}), and verification (\textsf{Verify}). The pseudo-codes for \textsf{KeyGen}, \textsf{Sign}, and \textsf{Verify} are illustrated in Algorithm \ref{algo-keygen}, \ref{algo-sign}, and \ref{algo-verify}. The polynomial ring in \textsf{Dilithium} is $\mathbb{Z}_q[x] /\left(x^{256}+1\right)$, $q = 8380417$. The core arithmetic operation in \textsf{Dilithium} is polynomial multiplication. \textsf{Dilithium} uses NTT to speed up polynomial multiplication.

\begin{algorithm}[!ht]
  \caption{\textsf{Dilithium}.$\mathsf{KeyGen}()$ \cite{ducas2018crystals}}
  \label{algo-keygen}
  \small
  \begin{algorithmic}[1]
    \Statex \textbf{Input:} $\zeta \leftarrow\{0,1\}^{256}$.
    \Statex \textbf{Output:} Public and secret keys $\left(p k=\left(\rho, \mathbf{t}_1\right), s k=\left(\rho, K, t r, \mathbf{s}_1, \mathbf{s}_2, \mathbf{t}_0\right)\right)$.
    \State $\left(\rho, \rho^{\prime}, K\right) \in\{0,1\}^{256} \times\{0,1\}^{512} \times\{0,1\}^{256}:=\mathrm{H}(\zeta)$ \Comment H is instantiated as \textsf{SHAKE-256}
    \State $\mathbf{A} \in \mathcal{R}_{q}^{k \times \ell} := \mathsf{ExpandA}(\rho)$ \Comment $\mathbf{A}$ is generated and stored in NTT Representation as $\hat{\mathbf{A}}$
    \State $\left(\mathbf{s}_1, \mathbf{s}_2\right) \in S_\eta^{\ell} \times S_n^k:=\mathsf{ExpandS}\left(\rho^{\prime}\right)$
    \State $\mathbf{t}:=\mathbf{A} \mathbf{s}_1+\mathbf{s}_2$ \Comment Compute $\mathbf{A s}_1$ as $\texttt{NTT}^{-1}\left(\hat{\mathbf{A}} \circ \texttt{NTT}\left(\mathbf{s}_1\right)\right)$
    \State $\mathbf{t}_{1} := \mathsf{Power2Round}_{q,d}(\mathbf{t})$
    \State $tr \in\{0,1\}^{256}:=\mathrm{H}\left(\rho \| \mathbf{t}_1\right)$
    \State \textbf{return} $\left(p k=\left(\rho, \mathbf{t}_1\right), s k=\left(\rho, K, t r, \mathbf{s}_1, \mathbf{s}_2, \mathbf{t}_0\right)\right)$
  \end{algorithmic}
\end{algorithm}

\begin{algorithm}[!ht]
  \caption{\textsf{Dilithium}.$\mathsf{Sign}(sk, M)$ \cite{ducas2018crystals}}
  \label{algo-sign}
  \small
  \begin{algorithmic}[1]
     \Statex \textbf{Input:} Secret key $s k=\left(\rho, K, t r, \mathbf{s}_1, \mathbf{s}_2, \mathbf{t}_0\right)$, Message $M \in \{0,1\}^*$.
     \Statex \textbf{Output:} Signature $\sigma=(\tilde{c}, \mathbf{z}, \mathbf{h})$.
    
    \State $\mathbf{A} \in \mathcal{R}_{q}^{k \times \ell} := \mathsf{ExpandA}(\rho)$ \Comment $\mathbf{A}$ is generated and stored in NTT Representation as $\hat{\mathbf{A}}$
    \State $\mu \in\{0,1\}^{512}:=\mathrm{H}(tr \| M)$
    \State $\kappa:=0,(\mathbf{z}, \mathbf{h}):=\perp$
    \State$\rho^{\prime} \in\{0,1\}^{512}:=\mathrm{H}(K \| \mu)$ (or $\rho^{\prime} \leftarrow\{0,1\}^{512}$ for randomized signing)
    
    \While {$(\mathbf{z}, \mathbf{h}) = \bot$} \Comment Precompute $\hat{\mathbf{s}}_1:=\texttt{NTT}\left(\mathbf{s}_1\right), \hat{\mathbf{s}}_2:=\texttt{NTT}\left(\mathbf{s}_2\right)$, and $\hat{\mathbf{t}}_0:=\texttt{NTT}\left(\mathbf{t}_0\right)$
        \State $\mathbf{y} \in \tilde{S}_{\gamma_1}^{\ell}:=\mathsf{ExpandMask}\left(\rho^{\prime}, \kappa\right)$
        \State $\mathbf{w} := \mathbf{Ay}$ \Comment $\mathbf{w}:=\operatorname{NTT}^{-1}(\hat{\mathbf{A}} \cdot \texttt{NTT}(\mathbf{y}))$
        \State $\mathbf{w}_1 := \mathsf{HighBits}_q(\mathbf{w}, 2\gamma_2)$
        \State $\tilde{c} \in\{0,1\}^{256}:=\mathrm{H}\left(\mu \| \mathbf{w}_1\right)$
        \State $c \in B_\tau:= \mathsf{SamplelnBall}(\tilde{c})$  \Comment{Store $c$ in NTT representation as $\hat{c}=\texttt{NTT}(c)$}
        \State $\mathbf{z} := \mathbf{y} + c\mathbf{s}_1$ \Comment{Compute $c \mathbf{s}_1$ as $\texttt{NTT}^{-1}\left(\hat{c} \cdot \hat{\mathbf{s}}_1\right)$}
        \State $\mathbf{r}_0:=\mathsf{LowBits}_q\left(\mathbf{w}-c \mathbf{s}_2, 2 \gamma_2\right)$ \Comment{Compute $c \mathbf{s}_2$ as $\texttt{NTT}^{-1}\left(\hat{c} \cdot \hat{\mathbf{s}}_2\right)$}
        \If {$||\mathbf{z}||_\infty \geq \gamma_1 - \beta$ or $||\mathbf{r}_0||_\infty \geq \gamma_2 - \beta$ }
            $(\mathbf{z}, \mathbf{h}) := \bot$
        \Else
            \State $\mathbf{h} := \mathsf{MakeHint}_q(-c\mathbf{t}_0, \mathbf{w} - c\mathbf{s}_2 + c\mathbf{t}_0, 2\gamma_2)$ \Comment{Compute $c \mathbf{t}_0$ as $\texttt{NTT}^{-1}\left(\hat{c} \cdot \hat{\mathbf{t}}_0\right)$}
            \If {$||c\mathbf{t}_0||_\infty \geq \gamma_2$ or $\mathsf{NumberOfOne}(\mathbf{h}) > \omega$}
                $(\mathbf{z}, \mathbf{h}) := \bot$
            \EndIf
        \EndIf
        \State $\kappa:=\kappa+\ell$
    \EndWhile

    \State \textbf{return} $\sigma=(\mathbf{z}, \mathbf{h}, c)$
  \end{algorithmic}
\end{algorithm}

\begin{algorithm}[!ht]
  \caption{\textsf{Dilithium}.$\mathsf{Verify}(pk, M, \sigma=(\tilde{c}, \mathbf{z}, \mathbf{h}))$ \cite{ducas2018crystals}}
  \label{algo-verify}
  \small
  \begin{algorithmic}[1]
     \Statex \textbf{Input:} Public key $pk = (\rho, \mathbf{t}_{1})$, Message $M \in \{0,1\}^*$, Signature $\sigma=(\tilde{c}, \mathbf{z}, \mathbf{h})$.
     \Statex \textbf{Output:} Result $r \in \{0,1\}$.
    
    \State $\mathbf{A} \in \mathcal{R}_{q}^{k \times \ell} := \mathsf{ExpandA}(\rho)$
    \State $\mu \in\{0,1\}^{512}:=\mathrm{H}\left(\mathrm{H}\left(\rho \| \mathbf{t}_1\right) \| M\right)$
    \State $c:=\mathsf{SamplelnBall}(\tilde{c})$
    \State $\mathbf{w}'_1 := \mathsf{UseHint}_q(\mathbf{h}, \mathbf{Az} - c\mathbf{t}_1 \cdot 2^d, 2\gamma_2)$
    
    \State \textbf{return} $\tilde{c}=\mathrm{H}\left(\mu \| \mathbf{w}_1^{\prime}\right)$ \textbf{and} $\|\mathbf{z}\|_{\infty}<\gamma_{1}-\beta$ \textbf{and} $\mathsf{NumberOfOne}(\mathbf{h}) \leq \omega$ 
  \end{algorithmic}
\end{algorithm}

\subsection{Functions}
We provide a brief introduction to various functions utilized in \textsf{Dilithium} here. \textsf{ExpandA} and \textsf{ExpandS} are sampling functions. \textsf{ExpandA} samples a matrix whose coefficients are in $\mathbb{Z}_q$ uniformly, so it can be assumed that the output of \textsf{ExpandA} is in the NTT domain. \textsf{ExpandS} generates vectors that coefficients within the interval $[-\eta,\eta]$. \textsf{SampleInBall} (Algorithm \ref{algo:sampleinball}) samples a polynomial with coefficients from $\{-1,0,1\}$ and only $\tau$ non-zero coefficients. The hash function $\mathrm{H}$ is instantiated as \textsf{SHAKE-256} \cite{dworkin2015sha}.
\begin{algorithm}[!ht]
 	\caption{\textsf{SampleInBall} $(\rho)$ \cite{ducas2018crystals}} \label{algo:sampleinball}
 	\begin{algorithmic}[1]
                \Statex Samples a polynomial $c \in R_q$ with coefficients from $\{-1,0,1\}$ and Hamming weight $\tau$.
 	          \Statex \textbf{Input:}  A seed $\rho \in\{0,1\}^{256}$.
 	        \Statex \textbf{Output:} A polynomial $c$ in $R_q$.
                \State  $c \leftarrow 0$ 
                \State  $k \leftarrow 8$ 
                \For {i = $256 - \tau$ to 255}
                \While{ $\mathrm{H}(\rho) \llbracket k \rrbracket > i $  }
                \State  $k \leftarrow k+1$ 
                \EndWhile
                \State  $j \leftarrow \mathrm{H}(\rho) \llbracket k \rrbracket$   \Comment{ $j$ is a pseudorandom byte that is $\leq i$}
                \State  $c_i \leftarrow c_j$ 
                \State  $c_j \leftarrow(-1)^{\mathrm{H}(\rho)[i+\tau-256]}$ 
                \State  $k \leftarrow k+1$ 
                \EndFor
                
                \Statex \textbf{return}   $c$
 	\end{algorithmic}
  \label{algo:sampleinball}
\end{algorithm}
\subsection{ARM Cortex-M4}
The ARM Cortex-M4 processor distinguishes itself as an efficient embedded processor built on the ARMv7E-M architecture. NIST chose ARM Cortex-M4 as its preferred microcontroller benchmark platform. The architecture of the ARM Cortex-M4 encompasses 16 32-bit general-purpose registers (GPRs), denoted as \texttt{r0-r15}. However, only 14 registers are available for general use. Additionally, the ARM Cortex-M4 integrates 32 floating-point register units (FPU) to temporarily store results from the GPRs. The ARM Cortex-M4 introduces advanced DSP instructions. DSP instructions enable simultaneous operations on 8-bit and 16-bit data elements within its GPRs. This distinctive capability facilitates the parallel processing of 4 and 2 elements. 


\subsection{Apple M2}
The Apple M2 processor operates on the AArch64 architecture, exclusive to ARMv8-A. ARMv8 is specifically designed for high-performance embedded applications, encompassing tablets (e.g., iPad Pro, iPad Air), laptops (e.g., MacBook Air, MacBook Pro), and mobile phones (e.g., iPhone, Samsung Galaxy). ARM supports Single Instruction Multiple Data (SIMD) instructions NEON. It comprises 32 128-bit vector registers, accommodating 8-bit, 16-bit, 32-bit, and 64-bit data element sizes. NEON supports Intrinsics to implement the NEON program conveniently.  Within NEON Intrinsics, various variable types are defined, including \texttt{int8x16\_t}, \texttt{int16x8\_t}, \texttt{int32x4\_t}, and \texttt{int64x2\_t}, representing data units processed by the function as 8 bits, 16 bits, 32 bits, and 64 bits, respectively.

\section{Sparse Polynomial Multiplication in Dilithium}
\label{sec-sparse}
\subsection{Index-based Sparse Polynomial Multiplication}
Polynomial multiplication is one of the computationally intensive operations in \textsf{Dilithium}. NTT is usually used to speed up polynomial multiplication in lattice-based cryptography. During the \textsf{Dilithium} signing process, there exists a challenge polynomial $c$. $c$ contains only $\tau$ positive or negative ones, while the remaining coefficients are zeros. For such sparse polynomials, polynomial multiplication can be achieved without resorting to NTT technology.  Zheng et al. \cite{zheng2022parallel} proposed an index-based sparse polynomial multiplication technique. As outlined in Algorithm \ref{algo-index}, coefficients of the challenge polynomial are scrutinized. Polynomial multiplication is streamlined into operations involving only addition, subtraction, and conditional checks. Compared to NTT, index-based sparse polynomial multiplication saves multiplication operations.

\subsection{Encode Challenge Polynomial}
In Algorithm \ref{algo-index}, a notable drawback is the presence of branch statements. Branch statements are time-consuming. To eliminate branches in sparse polynomial multiplication, we apply an encoding operation to the challenge polynomial $c$. Specifically, we introduce an index array named $index$, with a size of $\tau+1$. The first element of the array represents the count of positive indices in polynomial $c$, while the remaining $\tau$ elements encompass all continuous positive indices and continuous negative indices. The detailed steps of the \textsf{EncodeC} algorithm are outlined in Algorithm \ref{algo-encodec}.
\alglanguage{pseudocode}
\begin{algorithm}[h]
 	\caption{Index-based sparse multiplication \cite{zheng2022parallel}}
		\label{algo-index}
 	\begin{algorithmic}[1]
		\Statex \textbf{Input:}
			$\VT{c}=\SUM{i=0}{n-1}{c_i\cdot x^i}\in B_{\tau}, \ \VT{a}=\SUM{i=0}{n-1}{a_i\cdot x^i}\in \RingRq$.
		\Statex \textbf{Output:}
			$\VT{u}=\VT{c}\cdot \VT{a}\in \RingRq$.
		\For{$i\in \Array{0,1,\cdots,2n-1}$}
			\State	$w_i:=0$
		\EndFor
		\For{$i\in \Array{0,1,\cdots,n-1}$}
			\If{$c_i=1$}
				\For{$j\in \Array{0,1,\cdots, n-1}$}
					\State	$w_{i+j}:=w_{i+j}+a_{j}$
				\EndFor
			\EndIf
			\If{$c_i=-1$}
				\For{$j\in \Array{0,1,\cdots,n-1}$}
					\State	$w_{i+j}:=w_{i+j}-a_{j}$
				\EndFor
			\EndIf
		\EndFor
		\For{$i\in \Array{0,1,\cdots,n-1}$}
			\State	$u_i:=w_i - w_{i+n}\pmod{q}$
		\EndFor
		\State	$\VT{u}:=\SUM{i=0}{n-1}{u_i\cdot x^i}$
		\Comment{$\VT{u}\in \RingRq$}
		\Return $\VT{u}$
 	\end{algorithmic}
\end{algorithm}

 \alglanguage{pseudocode}
 
\begin{algorithm}[htbp]
 	\caption{Encode challenge polynomial $c$} \label{algo-encodec}
 	\begin{algorithmic}[1]
 	       \Statex \textbf{Input:} $c\in\mathbb{Z}_q[x]/(x^n+1)$, $\tau$ coefficients are in {-1,1}, the rest coefficients are 0.
 	        \Statex \textbf{Output:} Array $index$ of length $\tau+1$.
			
			\State $h=1,t=\tau$
			\For{$i\in  {0,\cdots, 255}$}
                    \If  {$c_i == 1$}
                    $index_{h} = i$, $h = h + 1$
                    \EndIf
                    \If  {$c_i == -1$}
                    $index_{t} = i$, $t = t - 1$
                    \EndIf
			\EndFor
                \State $index_{0}=h-1$
 		    \State \textbf{return}  Array $index$
 	\end{algorithmic}
\end{algorithm}

\subsection{Sparse Polynomial Multiplication without Branches}
Note that in the above \textsf{EncodeC} function, we can encode polynomial $c$ as an index array. Therefore, we can achieve sparse polynomial multiplication by directly traversing the index array without conditional branches. Based on this, we introduce a branchless sparse polynomial multiplication, as depicted in Algorithm \ref{algo-nobranch}.
\alglanguage{pseudocode}
\begin{algorithm}[h]
 	\caption{Sparse multiplication without branches}
		\label{algo-nobranch}
 	\begin{algorithmic}[1]
		\Statex \textbf{Input:}
			$\VT{a}=\SUM{i=0}{n-1}{a_i\cdot x^i}$, and index array $index$ of length $\tau+1$ containing the positive and negative ones, array $index$ = \textsf{EncodeC}$(c)$.
		\Statex \textbf{Output:}
			$\VT{u}=\VT{c}\cdot \VT{a}\in \RingRq$.
            \For{$i\in \Array{0,1,\cdots,2n-1}$}
            \State $w_i=0$
            \EndFor
            \State $poscnt \gets index_0$
		\For{$i\in \Array{1,2,\cdots,poscnt}$}
                \State $pos \gets index_{i}$
                \For{$j\in \Array{0,1,\cdots,n-1}$}
			\State	$w_{pos+j}=w_{pos+j}+a_{j}$
                \EndFor
		\EndFor
		\For{$i\in \Array{poscnt+1,2,\cdots,\tau}$}
                \State $pos \gets index_{i}$
                \For{$j\in \Array{0,1,\cdots,n-1}$}
			\State	$w_{pos+j}=w_{pos+j}-a_{j}$
                \EndFor
		\EndFor
		
		\For{$i\in \Array{0,1,\cdots,n-1}$}
			\State	$u_i=w_i - w_{i+n}\pmod{q}$
		\EndFor
		\State	$\VT{u}=\SUM{i=0}{n-1}{u_i\cdot x^i}$
		\Comment{$\VT{u}\in \RingRq$}
		\Return $\VT{u}$
 	\end{algorithmic}
\end{algorithm}

\section{Implementation}
\subsection{Discussion of Sparse Polynomial Multiplication Parallelism on 32-bit and 64-bit Platforms}
In this section, we delve into the parallelism aspects of sparse polynomial multiplication within the \textsf{Dilithium} scheme's signature process, specifically focusing on sparse polynomial vector multiplications denoted as $c\mathbf{s}_1$ and $c\mathbf{s}_2$. The corresponding norm values for $c\mathbf{s}_1$ and $c\mathbf{s}_2$ under three parameter sets are detailed in Table \ref{tab:parallelparams}.
\begin{table}[!ht]
\centering
\topcaption{Parameters of sparse polynomial multiplication.}
\begin{tabular}{ c|c|c|c }
\toprule
          & $\eta$ & $\tau$ & $||c\mathbf{s}_1||, ||c\mathbf{s}_2||$ \\ \hline
\textsf{Dilithium2} & 2   & 39  & 78      \\
\textsf{Dilithium3} & 4   & 49  & 196     \\ 
\textsf{Dilithium5} & 2   & 60  & 120     \\ \bottomrule
\end{tabular}

\label{tab:parallelparams}
\end{table}

For \textsf{Dilithium2} and \textsf{Dilithium5}, the coefficients of $c\mathbf{s}_1$ and $c\mathbf{s}_2$ fall within the range of 8-bit signed integers. This enables 4-way parallelism on ARM Corterx-M4 and 16-way parallelism on Apple M2 leveraging 128-bit ARM NEON vector registers. For \textsf{Dilithium3}, coefficients exceed the range of an 8-bit signed integer, thus hindering the feasibility of achieving 4 or 16-way parallelism. By outputting the coefficients of $c\mathbf{s}_1$ and $c\mathbf{s}_2$, we observe that, under the \textsf{Dilithium3} parameters, the majority of coefficient magnitudes in $c\mathbf{s}_1$ and $c\mathbf{s}_2$ are relatively small. Therefore, we conduct a probability analysis on $c\mathbf{s}_1$ and $c\mathbf{s}_2$ to assess the possibility of their coefficients falling within the range of an 8-bit signed integer (i.e., $[-128,127]$). Utilizing a probability distribution calculation based on the value distributions of polynomials $\mathbf{s}_1$ and $c$, our analysis indicates a probability of approximately $10^{-11}$, equivalent to a magnitude of $2^{-36}$. When utilizing 8-bit signed integer variables to store the coefficients of $c$, we can produce accurate signatures up to $2^{36}$ times, which is deemed acceptable for certain embedded IoT scenarios. It should be noted that when the number of signatures required is relatively small, the error rate using the above method is acceptable. But when the demand for signatures is large, we recommend still using classic NTT to calculate polynomial multiplication. To optimize parallelism, we propose a modified \textsf{Dilithium3} version designed to generate correct signatures up to $2^{36}$ times. The modified \textsf{Dilithium3} uses 8-bit signed integers to store sparse polynomial coefficients of \textsf{Dilithium3}.   The detailed possibility analysis is discussed as follows:

	 Let	$$
	c=\sum_{i=0}^{n-1} c_i x^i \in B_\tau, s=\sum_{i=0}^{n-1} s_i x^i \in R_q, u=c \cdot s=\sum_{i=0}^{n-1} u_i x^i \in R_q,
	$$
	
	We have $\forall i \in\{0,1, \cdots, n-1\}$ ,
	$$
	\begin{aligned}
		u_i & =\sum_{j=0}^i c_j s_{i-j}-\sum_{j=i+1}^{n-1} c_j s_{n+i-j} \\
		& =\sum_{k \in S} c_k s_{i-k}^{\prime} \\
		& =\sum_{k \in S} s_{i-k}^{\prime \prime}
	\end{aligned}
	$$
	
	among this $S=\left\{j \mid c_j \neq 0\right\} , \quad
	s_{i-k}^{\prime}=\left\{\begin{array}{l}s_{i-k}, i \geqslant k \\ -s_{n+i-k}, i<k\end{array}\right.$ ,
	$s_{i-k}^{\prime \prime}= \begin{cases}s_{i-k}^{\prime}, & c_k > 0 \\ -s_{i-k}^{\prime}, & c_k<0 \end{cases}$.
	
	Assuming a constant \(c\), where the coefficients of \(s_i\) follow a discrete uniform distribution on the interval \([- \eta, \eta]\) and are mutually independent, \(s_{i-k}^{\prime\prime}\) and \(s_i\) follow the same distribution. Consequently, \(u_i\) follows a discrete version of the Irwin-Hall distribution.

	$$
	\begin{aligned}
		p_i=P\left(u_k=i\right) & =P\left(s=\sum_{k=1}^\tau a_j=i\right) \\
		& =\frac{\sum_{s=i} 1}{(2 \eta +1)^\tau} .
	\end{aligned}
	$$
	$P\left(|u_k| > 128 \right) \approx 6.706350411547372 \times 10 ^ {-14}.$
	
	Let the random variable $X$ denote the number of coefficients in $cs$ that exceed their specified bounds. Assuming each coefficient is independent of others, $X \sim B(n, p)$, where $n$ is the number of coefficients and $p$ is the probability of a coefficient exceeding the bound. Therefore, the probability that at least one coefficient in $u_i$ exceeds 8 bits signed integer is given by
\[
\begin{aligned}
	P\left(X \geq 1 \right) & = 1 - P\left(X = 0\right) \\
	& = 1 - (1 - p)^{256} \\
	& \approx 1.716671249596402 \times 10^{-11}.
\end{aligned}
\]

\subsection{Sparse Polynomial Multiplication on ARM Cortex-M4}

In this section, we discuss our implementation approaches for polynomial multiplication in \textsf{Dilithium} on ARM Cortex-M4. Parallel sparse polynomial multiplication presented by Zheng et al. \cite{zheng2022parallel} is not suitable for ARM Cortex-M4 implementation. It necessitates additional table storage for concatenated polynomial vectors and is unfriendly for resource-constrained platforms. In our work, we implement Algorithm \ref{algo-nobranch} and opt to employ 8-bit signed arrays to store vectors $\mathbf{s}_1$ and $\mathbf{s}_2$, as well as polynomial  $c$. 

We present the implementation of Algorithm \ref{algo-nobranch}. Our implementation is depicted in Algorithm \ref{algo:sparsem4v2} in Appendix \ref{appendix:sparsem4ap} . We compute \(\texttt{ret}[i] = \texttt{ret}[i] + c[k] \cdot s[i - k]\), where \(k \in \left\{ k | c[k] \neq 0 \right\}\). This reduces the number of accesses to the \texttt{ret} array to \(n\), saving \((\tau - 1) \cdot n\) accesses. Additionally, to avoid the need for subtraction at the end, we preprocess the \texttt{s} array from \(\texttt{s} = (s_0, s_1, ..., s_{n-1})\) to \((-s_0, -s_1, ..., -s_{n-1}, s_0, s_1, ..., s_{n-1})\). This changes the processing of \texttt{ret} coefficients to \(\texttt{ret}[i] = \texttt{ret}[i] + c[k] \cdot s[n + i - k]\). Besides, the resulting array is also of length \(n\).                              We employ two techniques to fully leverage the pipeline. Firstly, we implement loop unrolling using assembly pseudo-instructions  \texttt{.rept} and \texttt{.endr}. With loop unrolling, we avoid the use of conditional instructions, preventing pipeline stalls. Secondly, we load as many coefficients as possible in a single iteration. We store values temporarily in FPU and then retrieve them back to registers after the loop completes. We utilize \texttt{ldmia} and \texttt{stmia} to load and store in a parallel way.

\paragraph{Adjustments for Signing Procedure Optimization.}

The above-mentioned 
precomputation-based sparse polynomial multiplication 
requires modifications on the input polynomial $s$. However, rejection sampling needs to be performed in the signing procedure to generate valid signatures. $\mathbf{s}_1$ and $\mathbf{s}_2$ remain unchanged after a restart. Therefore, we make some adjustments in sparse polynomial multiplication implementation. Specifically, we encapsulate the preprocessing of $s$ into \textsf{skDecode}. We retain the remaining code excluding preprocessing steps in Algorithm \ref{algo:sparsem4v2}. In doing so, we ensure that $\mathbf{s}_1$ and $\mathbf{s}_2$ will not be modified in each restart of the signing process within the loop. Since the input $s$ for sparse polynomial multiplication is a $2n$-size 8-bit signed integers array, we make corresponding modifications to the private key decoding function (\textsf{skDecode}). We decode the private key to a $2n$-size array, with the first $n$ elements corresponding to the coefficients of $\mathbf{s}_1$ and $\mathbf{s}_2$, and the last $n$ elements being zero. We implement modified  \textsf{skDecode} using DSP instructions for ARM Cortex-M4. This is feasible since the 32-bit GPRs of ARM Cortex-M4 can store 4 8-bit integers, allowing for a 4-way parallel acceleration of the private key decoding process. In addition, for the rejection process in the signature, we use early rejection after each polynomial is calculated since it can reduce unnecessary calculations. If the rejection condition is satisfied, it will be rejected immediately and restarted. Otherwise, execution will continue. As stated in \cite{zheng2023optimized}, the probability of vector $\mathbf{z}$ falling within a good range is always greater than the probability of vector $\mathbf{r}_0$. Therefore, we check vector $\mathbf{r}_0$ first and then vector $\mathbf{z}$ to speed up the signing process.

\subsection{Sparse Polynomial Multiplication on Apple M2}

In this section, we discuss how to implement sparse polynomial multiplication on Apple M2 to achieve better performance.
The Apple M2 utilizes the ARMv8-A architecture and supports ARM NEON instruction sets. We employ ARM NEON Intrinsics for vectorizing sparse polynomial multiplication. Specifically, in Algorithm \ref{algo-nobranch}, polynomial $a$ needs to be iterated for cumulative addition and subtraction operations. We identify this segment as suitable for vectorization. We efficiently load the polynomial $a$ in parallel into ARM NEON Intrinsics-defined vector register variables (\texttt{int8x16\_t}), allowing for 16 simultaneous operations using the \texttt{vld1q\_s8} function.

Moreover, we implement an early rejection technique on vector registers. Specifically, during the calculation of $c\mathbf{s}_1$ and $c\mathbf{s}_2$, we simultaneously include the computations for $\mathbf{y}+c\mathbf{s}_1$ and $\mathbf{w}_0-c\mathbf{s}_2$. In previous implementations, the typical approach involves computing the entire vector result for $c\mathbf{s}_1$ and then adding the coefficients of vector $\mathbf{y}$ to obtain the entire vector result for $\mathbf{y}+c\mathbf{s}_1$. Subsequently, an infinity norm check is performed to determine if any coefficients exceed a predefined threshold. However, this approach leads to redundant computations. For example, if the first coefficient of the first-dimensional vector of $\mathbf{y}+c\mathbf{s}_1$ encounters an issue, all subsequent coefficient calculations are rendered invalid, necessitating a signature restart.

To address this, we fuse the computations of $\mathbf{y}+c\mathbf{s}_1$ and $\mathbf{w}_0-c\mathbf{s}_2$ into the sparse polynomial multiplication implementation. Each vectorized calculation computes results for 16 coefficients. As these coefficients reside in a single vector register, we employ the \texttt{vcgeq\_s32} function to compare values in two vector registers, yielding a 128-bit mask register with comparison results for each 32-bit integer. The \texttt{vaddvq\_u32} instruction is then used to sum all 32-bit data units in the mask vector register. If the sum is non-zero, it indicates the presence of illegal data in the current computation, prompting an immediate return of 1 to restart the signature.

In addition to the infinity norm check, the addition process of $\mathbf{y}+c\mathbf{s}_1$ involves the addition of 8-bit and 32-bit integers. We address this by first extending the 8-bit integers in the 128-bit vector register into 32 bits, employing the \texttt{vmovl\_s16} and \texttt{vmovl\_s8} functions. Since direct extension from 8 to 32 bits is not supported in ARM NEON, we first extend to 16 bits and then to 32 bits. This vectorized approach, coupled with early checks, significantly enhances the efficiency of calculations involving $\mathbf{y}+c\mathbf{s}_1$ and $\mathbf{w}_0-c\mathbf{s}_2$  in the signature process.

\subsection{Other Optimization Modules}
In addition to sparse polynomial multiplication, we optimize other modules that contribute to key generation and verification procedure performance on the ARM Cortex-M4 platform. We implement assembly for rejection sampling, polynomial addition, and polynomial subtraction, which have not been done in ARM assembly before. Specifically, we find that the previous implementation has assembly for rejection sampling of coefficients in $\mathbb{Z}_q$ but does not have assembly implementations for sampling coefficients in the ranges \([- \eta, \eta]\) and \([- (\gamma_1-1), \gamma_1]\). Since the sampling approach for these polynomial coefficients is rejection sampling, the implementation ideas are similar. We implement ARM assembly for \texttt{rej\_eta} and \texttt{rej\_gamma1}, corresponding to polynomials sampled in the ranges \([- \eta, \eta]\) and \([- (\gamma_1-1), \gamma_1]\), respectively. We utilize the \texttt{.rept} pseudo-instruction for loop unrolling to reduce the number of conditional branches and jumps. The optimized \texttt{rej\_eta} achieves a speedup of 42.3\% for \(\eta = 2\) and 33.3\% for \(\eta = 4\) compared to the previous implementation. Additionally, we implement ARM assembly for encoding secret polynomial vectors and decoding private keys. In our implementation, $\mathbf{t}_1$ and $\mathbf{t}_0$ use 16-bit signed integer arrays since their coefficients fall within the 16-bit signed integer range. This allows for ARM assembly parallelization during the encoding of $\mathbf{t}_1$ and $\mathbf{t}_0$, resulting in a 30\% - 50\% speedup compared to the scalar C implementation. For the decoding of private keys during the signing process, as we employ sparse polynomial multiplication, secret vectors $\mathbf{s}_1$ and $\mathbf{s}_2$ need to be decoded into 8-bit signed integer arrays. This process can be efficiently implemented in ARM assembly with a 4-way parallelization.

\section{Results}
In the preceding sections, we introduce optimization techniques for \textsf{Dilithium} on ARM Cortex-M4 and Apple M2. In this section, we present the performance evaluation brought about by these optimization techniques on the two platforms. We evaluated our ARM Cortex-M4 implementation on \texttt{STM32F407G-DISC1} development board within the \texttt{pqm4} framework \cite{2019pqm4}. We used \texttt{gcc version 14.0.3} with the \texttt{-O3}. We evaluated our Apple M2 implementation on MacBook Air 2022. We used \texttt{Clang version 12.0.4} with \texttt{-O3}. Our \textsf{Dilithium} ARM Cortex-M4 implementation was built upon the open-source ARM Cortex-M4 code in \cite{huang2024revisiting} \footnote{\url{https://github.com/UIC-ESLAS/Dilithium-Multi-Moduli/tree/master/M4}}. Our Apple M2 implementation was built upon the \textsf{Dilithium} ARM NEON open-source code \footnote{\url{https://github.com/neon-ntt/neon-ntt}}. It should be noted that our testing platform was different from related work in \cite{huang2024revisiting}, which used \texttt{STM32F407VG}. Our test board was \texttt{STM32F407G-DISC1}. Therefore, we also benchmarked the code in \cite{huang2024revisiting} for comparison.
\subsection{Sparse Polynomial Multiplication Results}
We compared our optimization techniques with the latest implementations of the \textsf{Dilithium} scheme on ARM Cortex-M4 \cite{huang2024revisiting} and Apple M2 \cite{2021Neon}. 

\begin{figure}[htbp]
  \centering
  \includegraphics[width=0.5\textwidth]{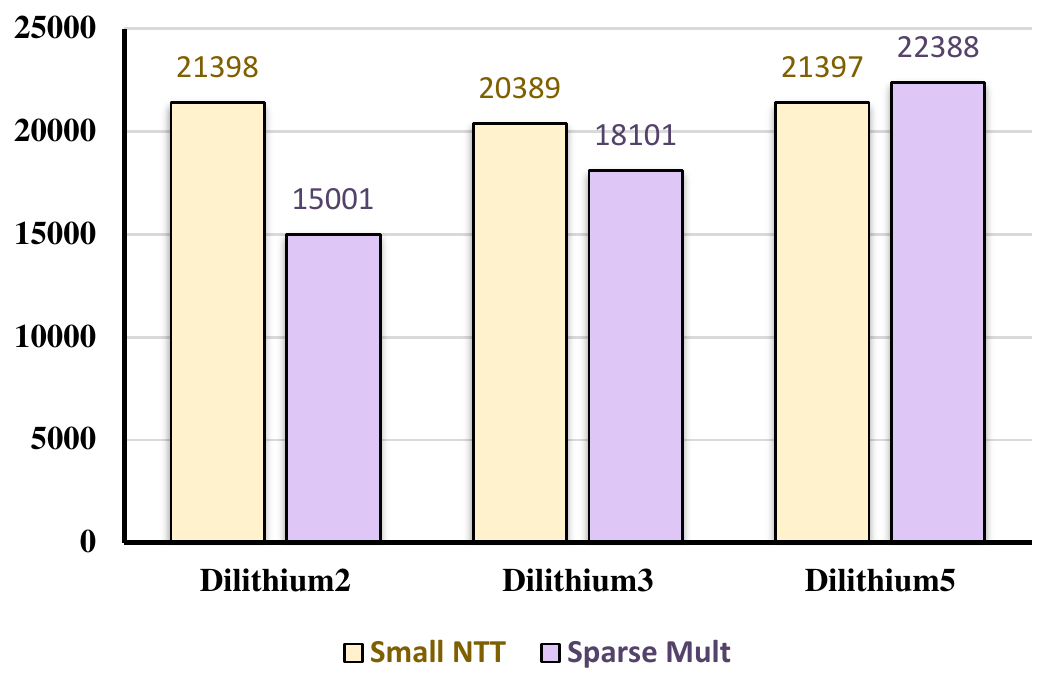}
  \caption{Polynomial multiplication evaluation on Cortex-M4.}
  \label{fig:sparsem4}
\end{figure}

Figure \ref{fig:sparsem4} presents the clock cycle counts for single polynomial multiplication using different implementation methods.  We obtain speedups in \textsf{Dilithium2} and \textsf{Dilithium3} parameters of 30\% and 11\% compared to small-modulus NTT in \cite{huang2024revisiting}. However, in \textsf{Dilithium5}, our implementation is slightly slower than the small-modulus NTT in \cite{huang2024revisiting}. 

Table \ref{tab:sparsea72} presents a comparison of $c\mathbf{s}_1$ and $c\mathbf{s}_2$ on the Apple M2 using NTT. The sparse polynomial multiplication technique achieves a performance improvement ranging from 33\% to 55\% compared to NTT technology in polynomial vector multiplication. The key factor contributing to this enhancement is the adoption of sparse polynomial multiplication, which utilizes 8-bit storage for polynomial coefficients, achieving 16-way parallelism under ARM NEON 128-bit vector registers.



\begin{table}[htbp]
\centering
\topcaption{CPU cycle counts comparison for $c\mathbf{s}_1$, $c\mathbf{s}_2$ using different techniques multiplication on Apple M2.}
\begin{tabular}{cccc}
\toprule
            & \textsf{Dilithium2}   & \textsf{Dilithium3}   & \textsf{Dilithium5}   \\ \midrule

$c\mathbf{s}_1$ (NTT \cite{2021Neon})    & 5270        & 6367        & 8761        \\ \hline
$c\mathbf{s}_1$ (Sparse) & 2358  & 4294  & 5675   \\ \hline
$c\mathbf{s}_2$ (NTT \cite{2021Neon})   & 5270        & 7652        & 9898        \\ \hline
$c\mathbf{s}_2$ (Sparse) & 2358   & 4083   & 6452 \\ \bottomrule

\end{tabular}
\label{tab:sparsea72}
\end{table}



\subsection{Scheme Results}
As shown in Table \ref{tab:schemespeedstack}, we present the performance test results of \textsf{Dilithium}. Compared to \cite{huang2024revisiting}, we achieve performance improvements of 2.0\%-3.2\%, 0.4\%-0.8\%, and 0.4\%-0.7\% in \textsf{KeyGen}, \textsf{Sign}, and \textsf{Verify}, respectively. The primary source of performance enhancement lies in the optimized sparse polynomial multiplication implementation in \textsf{Sign}.  Besides, we reduce stack usage by 1.2\% to 10.8\% in the signing procedure compared to \cite{huang2024revisiting}. It also demonstrates that sparse polynomial multiplication conserves resources more effectively compared to small-modulus NTT.

\begin{table}[htbp]
\centering
\topcaption{Speed and stack results comparison over 10000 iterations for \textsf{Dilithium} on ARM Cortex-M4.}
\begin{tabular}{@{}ccccccc@{}}
\toprule
Procedure               & Scheme                      & Work                                        & Speed{[}cc{]} & Stack{[}B{]} \\ \hline
\multirow{6}{*}{\textsf{KeyGen}} & \multirow{2}{*}{\textsf{Dilithium2}} & \cite{huang2024revisiting} & 1431405                 & 8516                    \\
                        &                             & Our work                             & 1385570                 & 8516                    \\
                        & \multirow{2}{*}{\textsf{Dilithium3}} & \cite{huang2024revisiting} & 2418966                 & 9548                    \\
                        &                             & Our work                             & 2367016                 & 9548                    \\
                        & \multirow{2}{*}{\textsf{Dilithium5}} & \cite{huang2024revisiting} & 4077258                 & 11596                   \\
                        &                             & Our work                             & 3994069                 & 11596                   \\ \hline
\multirow{6}{*}{\textsf{Sign}}   & \multirow{2}{*}{\textsf{Dilithium2}} & \cite{huang2024revisiting} & 3637132                 & 49372                   \\
                        &                             & Our work                             & 3608493                 & 44044                   \\
                        & \multirow{2}{*}{\textsf{Dilithium3}} & \cite{huang2024revisiting} & 5891490                 & 68932                   \\
                        &                             & Our work                             & 5866883                 & 68108                   \\
                        & \multirow{2}{*}{\textsf{Dilithium5}} & \cite{huang2024revisiting} & 7720252                 & 115924                  \\
                        &                             & Our work                             & 7853055                 & 107044                  \\ \hline
\multirow{6}{*}{\textsf{Verify}} & \multirow{2}{*}{\textsf{Dilithium2}} & \cite{huang2024revisiting} & 1428063                 & 8884                   \\
                        &                             & Our work                             & 1418269                 & 8884                   \\
                        & \multirow{2}{*}{\textsf{Dilithium3}} & \cite{huang2024revisiting} & 2325433                 & 9844                   \\
                        &                             & Our work                             & 2310862                 & 9844                   \\
                        & \multirow{2}{*}{\textsf{Dilithium5}} & \cite{huang2024revisiting} & 3999194                 & 11892                   \\
                        &                             & Our work                            & 3982393                 & 11892                   \\ 
                         \bottomrule
\end{tabular}
\label{tab:schemespeedstack}
\end{table}
\begin{table}[!ht]
\centering
\topcaption{Sign CPU cycles comparison on Apple M2 over 1000 iterations.}
\begin{tabular}{@{}ccc@{}}
\toprule
           & Sign \cite{2021Neon}   & Sign (Our Work) \\ \midrule
\textsf{Dilithium2} & 224666 & 201449   \\
\textsf{Dilithium3} & 355583 & 316875   \\
\textsf{Dilithium5} & 420376 & 375860   \\ \bottomrule
\end{tabular}
\label{tab:schemeapplem2}
\end{table}
On Apple M2, our optimization efforts are exclusively focused on enhancing the signing process. As a result, the provided data in Table \ref{tab:schemeapplem2} specifically compares results related to the signature process. Our optimized implementation utilizing sparse polynomial multiplication achieves an improvement of 10\% to 11\% in the signing performance of the scheme.

\section{Conclusion}
In this paper, we present efficient sparse polynomial multiplication implementation of the \textsf{Dilithium} scheme on both Cortex-M4 and Apple M2 platforms. We enhance the signature process by improving sparse polynomial multiplication, encoding sparse polynomials, and storing them based on positive and negative indices. We introduce a branchless method for sparse polynomial multiplication. We discuss the parallelization of sparse polynomial multiplication on embedded platforms. On ARM Cortex-M4, we utilize DSP instructions to accelerate sparse polynomial multiplication. We also optimize some small modules of \textsf{Dilithium}. On the Apple M2 platform, we leverage ARM NEON Intrinsics vectorization to implement sparse polynomial multiplication. This approach outperforms the latest ARM NEON implementation using NTT. We set a new speed record for \textsf{Dilithium} on ARM Cortex-M4 and Apple M2 platforms. Our implementation can be extended to other lattice-based cryptographic schemes involving sparse polynomial multiplication and applied to ARM 32-bit and 64-bit processors based on ARMv7 and ARMv8 architectures.

%
%
%
\bibliographystyle{splncs04}
%




\bibliography{sample-base}
\clearpage
\appendix
\section{Sparse Polynomial Multiplication ARM Cortex-M4 Implementation}
\label{appendix:sparsem4ap}
\begin{algorithm}[!htbp]
 	\caption{Pseduo-code for sparse polynomial multiplication using precomputation with ARMv7-m instructions} \label{algo:sparsem4v2}
 	\begin{algorithmic}[1]
 	       \Statex \textbf{Input:}  \texttt{ptr\_index}, the pointer of \texttt{uint8\_t pos\_neg\_list[$\tau$ + 1]}, \texttt{ptr\_s}, the pointer of \texttt{int8\_t s[$2n$]}.
 	        \Statex \textbf{Output:} \texttt{ptr\_ret}, the pointer of \texttt{int8\_t ret[$n$]}.
          
                \State \texttt{ldrb pos\_cnt, [ptr\_index], \#1} 
                \State \texttt{rsb.w neg\_cnt, pos\_cnt, $\tau$} 
                \State \texttt{add.w ptr\_s, \#256} \Comment pointer move to $s_{n + i}$, where $i$ initialized with 0 
                \State \texttt{.rept 16}
                \State \texttt{ldmia.w ptr\_ret, $\{$r0, r1, r2, r3$\}$}
                \State 1: \Comment{process the positive indices}
                \State \texttt{ldrb pos, [ptr\_index],\#1} \Comment Compute the number of positive indexes 
                \State \texttt{sub.w r\_pos, ptr\_s, pos} \Comment{\texttt{r\_pos} is the temp pointer to store offset of $s_{n + i -pos}$}
                \State \texttt{ldmia r\_pos!, $\{$r4, r5, r6, r7$\}$}
		    \State \texttt{sadd8 r0, r0, r4}
                \State \texttt{sadd8 r1, r1, r5} \Comment{(\texttt{r0}, \texttt{r1}, \texttt{r2}, \texttt{r3}) += (\texttt{r4}, \texttt{r5}, \texttt{r6}, \texttt{r7})}
                \State \texttt{sadd8 r2, r2, r6}
                \State \texttt{sadd8 r3, r3, r7}
                \State \texttt{subs.w pos\_cnt, \#1}
                \State \texttt{bne.w 1b}
                \State \texttt{stmia ptr\_ret!, $\{$r0, r1, r2, r3$\}$}
                \State \texttt{vmov pos\_cnt, r4}
                \State \texttt{sub.w ptr\_index, pos\_cnt}
                \State \texttt{add.w ptr\_s, \#16}
                \State \texttt{.endr}
                \State \texttt{.rept 16}  
                \State \texttt{ldmia.w ptr\_ret,$\{$r0, r1, r2, r3$\}$} 
                \State \texttt{2:} \Comment{process the negative indices}
                \State \texttt{ldrb pos, [ptr\_index], \#1} 
                \State \texttt{sub.w r\_pos, ptr\_s, pos} 
                \State \texttt{ldmia.w r\_pos,  $\{$r4, r5, r6, r7$\}$} 
                \State \texttt{ssub8 r0, r0, r4} 
                \State \texttt{ssub8 r1, r1, r5} \Comment{(\texttt{r0}, \texttt{r1}, \texttt{r2}, \texttt{r3}) -= (\texttt{r4}, \texttt{r5}, \texttt{r6}, \texttt{r7})}
                \State \texttt{ssub8 r2, r2, r6} 
                \State \texttt{ssub8 r3, r3, r7} 
                \State \texttt{subs.w neg\_cnt, \#1} 
                \State \texttt{bne.w 2b} 
                \State \texttt{stmia ptr\_ret!, $\{$r0, r1, r2, r3$\}$}  
                \State \texttt{vmov neg\_cnt, r5}
                \State \texttt{sub.w ptr\_index, neg\_cnt}
                \State \texttt{add.w ptr\_s, \#16} \Comment{each iteration 16 coefficients are processed, so $i$ += 16}
                \State \texttt{.endr}

                \Statex \textbf{return}  \texttt{ptr\_ret}
 	\end{algorithmic}
  \label{algo:sparsem4v2}
\end{algorithm}
\end{sloppypar}
\end{document}